\documentstyle[12pt]{article}
\begin{document}
	\title{Description of Nuclear Structure Effects in Subbarrier
Fusion by the Interacting Boson Model}
	\author{A.B. Balantekin\\
Department of Physics, University of Wisconsin,\\
Madison, Wisconsin 53706 USA\\
and\\
S.  Kuyucak\\ Department of Theoretical Physics, Research School
of Physical Sciences,\\
Australian National University, Canberra, ACT 0200 Australia}

\date{}
\maketitle

\begin{abstract}
Recent theoretical developments in using the Interacting Boson Model to 
describe nuclear structure effects in fusion reactions below the Coulomb 
barrier are reviewed.  Methods dealing with linear and all orders coupling 
between the nuclear excitations and the translational motion are discussed, 
and the latter is found to lead to a better description of the barrier 
distribution data.  A systematic study of the available data (cross 
sections, barrier and spin distributions) in rare-earth nuclei is 
presented.
\end{abstract}

\section{Introduction}

\indent
It is by now well established that to describe fusion cross sections
one  needs to include coupling of relative motion to other degrees of
freedom  \cite{st,yenibaha}.  These extra degrees of freedom yield a
distribution of barriers \cite{neil} and consequently enhance the
cross  section below the barrier \cite{ap,hen}.   Enhancement of
fusion cross  sections due to coupling of levels in colliding nuclei
to the relative  motion, has opened a new avenue for testing nuclear
structure models.  The  recent experimental determination of average
angular momenta \cite{van} and  the barrier distributions \cite{jack}
have provided even more stringent  tests for such models.

The natural language to study multidimensional barrier penetration is
the coupled channels formalism, which was widely used in investigating
subbarrier fusion phenomena \cite{st}.  An alternative formulation of
the multidimensional quantum tunneling is given by the path integral
formalism \cite{ap}. An algebraic nuclear structure model
significantly simplifies evaluation of the path integral.  The
Interacting Boson Model (IBM) of Arima and Iachello \cite{ibm} is one
such model which has been successfully employed to describe the
properties of low-lying collective states in medium heavy nuclei. Here
attempts to use IBM in describing nuclear structure effects in
subbarrier fusion are reviewed.

Path integral formulation of this problem, as sketched in the next
section, requires analytic solutions for the nuclear wave
functions. In our first attempt in using IBM to describe nuclear
structure effects in subbarrier fusion, we employed the SU(3) limit of
IBM \cite{ibm1}. However, the SU(3) limit corresponds to a rigid
nucleus with a particular quadrupole deformation and no hexadecapole
deformation, a situation which is not realized in most deformed
nuclei. Thus analytic solutions away from the limiting symmetries of
the IBM are needed for realistic calculations of subbarrier fusion
cross sections.  In a parallel development, a 1/N expansion was
investigated \cite{serdar1} for the IBM which provided analytic
solutions for a general Hamiltonian with arbitrary kinds of
bosons. This technique proved useful in a variety of nuclear structure
problems where direct numerical calculations are prohibitively
difficult. As we briefly describe in the next section, using the 1/N
expansion technique in the path integral formulation of the fusion
problem \cite{ibm2} makes it possible to go away from the three
symmetry limits of IBM, in particular, arbitrary quadrupole and
hexadecapole couplings can be introduced.

Recently a series of recent high-precision measurements were carried
out at the Australian National University (ANU) where distributions of
fusion barriers were determined directly from the fusion data
\cite{jack}.  IBM based analysis of fusion cross sections describes
the ANU data well \cite{ibm2,ibm3,ibm4,ibm5}, especially when the
higher order coupling effects are included.

In the next section, first the influence functional method is
summarized and its application to the linear coupling of translational
motion to the structure of target nuclei is described. The
significance of nonlinear couplings is discussed next, and a Green's
function method is introduced to describe such nonlinear effects.
Section 3 includes attempts to describe subbarrier fusion data with
these techniques.

\section{Algebraic Models in Subbarrier Fusion}

\indent
The Hamiltonian for the multi-dimensional quantum tunneling problem
relevant to subbarrier fusion is
\begin{equation}
H= -\frac{\hbar^2}{2\mu}\nabla^2 + H_{0}(\xi) + H_{\rm int} ({\bf r},\xi)
\end{equation}
where ${\bf r}$ is the relative coordinate of the target and
projectile, and ${\xi}$ represents any internal degrees of freedom of
the target.  $H_0 (\xi)$ and $H_{\rm int} ({\bf r},\xi)$ describe the
internal structure of the target nucleus and its coupling to the
relative motion, respectively.  In Eq.~(1), the propagator to go from
an initial state characterized by the relative radial coordinate $r_i$
(the magnitude of $\bf r$) and internal quantum numbers $n_i$ to a
final state characterized by $r_f$ and $n_f$ may be written as
\begin{equation}
K(r_f,n_f,T;r_i,n_i,0)=\int{\cal D}\left[r(t)\right]
e^{iS(r,T)/\hbar} \langle n_f |
\hat U_{\rm int} (r(t),T) |n_i \rangle,
\end{equation}
where $S(r,T)$ is the action for the translational motion and  $\hat
		      U_{\rm int}$ satisfies the differential equation
\begin{eqnarray}
    i\hbar\frac{\partial\hat U_{\rm int}}{\partial t} &=& \left[ H_0 +
  H_{\rm int} \right] \hat U_{\rm int},\\ \hat U_{\rm int}(t=0) &=& 1.
\end{eqnarray}
We want to consider the case where $r_i$ and $r_f$ are on opposite
sides of the barrier. In the limit when the initial and final states
are far away from the barrier, the transition amplitude is given by
the $S$-matrix element, which can be expressed in terms of the
propagator as \cite{ap}
\begin{eqnarray}
S_{n_f,n_i}(E) &=& \frac{i}{\hbar}\lim_{r_i\rightarrow\infty\atop
r_f\rightarrow-\infty}
\left(\frac{p_ip_f}{\mu^2}\right)^{1/2} 
\exp\left[i(p_fr_f-p_ir_i)/\hbar\right]\nonumber\\ &&
\times \int\limits_0^\infty dT e^{+iET/\hbar}K(r_f,n_f,T;r_i,n_i,0),
\end{eqnarray}
where $p_i$ and $p_f$ are the classical momenta associated with $r_i$
and $r_f$. In heavy ion fusion we are interested in the transition
probability in which the internal system emerges in any final
state. For the $\ell$th  partial wave, this is
\begin{equation}
T_\ell=\sum_{n_f}|S_{n_f,n_i}(E)|^2,
\end{equation}
which becomes, upon substituting Eqs.\ (2) and (5),
\begin{eqnarray}
T_\ell &=& \lim_{r_i\rightarrow\infty\atop r_f\rightarrow-\infty}
\left(\frac{p_ip_f}{\mu^2}\right)\int\limits_0^\infty
dTe^{iET/\hbar}\int\limits_0^\infty\tilde
Te^{-iE\tilde T/\hbar}\nonumber\\ 
&&\int{\cal D}[r(t)]\int{\cal D}[\tilde r(\tilde t)] 
e^{i[S(r,T)-S(\tilde r,\tilde T)]/\hbar}
\rho_M(\tilde r(\tilde t),\tilde T; r(t),T).
\end{eqnarray}
Here we have assumed that the energy dissipated to the internal system
is small compared to the total energy and taken $p_f$ outside the sum
over final states.  We have also identified the two-time influence
functional in Eq.~(7) as
\begin{equation}
\rho_M(\tilde r(\tilde t),\tilde T;r(t),T) =\langle n_i|
\hat U_{\rm int}^{\dagger} ({\tilde r}({\tilde t}),{\tilde T}) \hat
U_{\rm int} (r(t), T)| n_i\rangle,
\end{equation}
where the completeness of final states was used.  Eq.~(8) displays the
utility of the influence functional method when the internal system
has symmetry properties. If the Hamiltonian in Eq.~(3) has a dynamical
or spectrum generating symmetry, i.e. if it can be written in terms of
the Casimir operators or generators of a given Lie algebra, then the
solution of Eq.~(3) is an element of the corresponding Lie group
\cite{ap}.  Consequently the two time influence functional of Eq.~(8)
is simply a diagonal group matrix element for the lowest-weight state
and it can be evaluated using standard group-theoretical methods.

If the Interacting Boson Model is used to describe the effects of
nuclear structure on fusion, $H_0=H_{\rm IBM}$, and there are two
choices for $H_{\rm int}$.  One possibility is to take $H_{\rm int}$
to be of the form of the most general one-body transition operator for
IBM
\begin{equation}
H_{\rm int}= V(r) +\sum_{kjl} \alpha_{kjl}(r) [b^\dagger_j\tilde
b_l]^{(k)}\cdot Y^{(k)}(\hat{\bf r}),
\end{equation}
where  $b^\dagger_j$ and $b_l$ denote the boson creation and 
annihilation operators ($b_0=s$, $b_2=d$, etc).
The $k$ sum runs over $k=2,4,\ldots 2l_{\max}$, with $l_{\max}$
representing the maximum spin of boson operators. Odd values of $k$
are excluded as a consequence of the reflection symmetry of the
nuclear shape and the $k=0$ term is already included in the bare
potential $V(r)$. The form factors $\alpha_{kjl}(r)$ represent the
spatial dependence of the coupling between the intrinsic and
translational motions. The interaction term given in Eq.~(9) is an
element of the SU(6) algebra for the original form of the
IBM with $s$ and $d$ bosons, and is an element of the
SU(15) algebra when $g$ bosons are included as well \cite{ibm}.

To simplify the calculation of the influence functional, we can
perform a rotation at each instant to a frame in which the $z$-axis
points along the direction of relative motion. Neglecting the
resulting centrifugal and Coriolis terms in this rotating frame is
equivalent to ignoring the angular dependence of the original
Hamiltonian.  In this approximation, the coupling form factors become
independent of $\ell$ and only $m=0$ magnetic sub-states of the target
are excited \cite{nob}. For heavy systems, the neglected centrifugal
and Coriolis forces are small. We take the scattering to be in the
$x$-$y$ plane. Then making a rotation through the Euler angles
$\hat{\bf b}=(\phi,\pi/2,0)$, we can write the Hamiltonian as the
rotation of a simpler Hamiltonian depending only on the magnitude of
$\bf r$
\begin{equation}
H=R(\hat{\bf b})H^{(0)}(r)R^\dagger(\hat{\bf b}).\label{ham}
\end{equation}
Since in Eq.~(1) $H_0$ and $H_k$ are rotationally invariant, $H_{\rm
int}$ is the only term whose form is affected by this
transformation. Hence we introduce the rotated interaction Hamiltonian
$H_{\rm int}^{(0)}(r)$, given by
\begin{eqnarray}
H_{\rm int} &=& R(\hat{\bf b})H_{\rm int}^{(0)}(r)R^\dagger(\hat{\bf b}),\\ 
H_{\rm int}^{(0)}(r) &=& \sum_{jl m}\phi_{jl
m}(r)b^\dagger_{jm}b_{l m},\\ 
\phi_{jl m}(r) &=&
(-)^m\sum_k\sqrt{\frac{2k+1}{4\pi}}\langle jml
-m|k0\rangle\alpha_{kjl}(r).
\end{eqnarray}
If we assume now that the form factors $\alpha_{kjl}(r)$ are all
proportional to the same function of $r$, then the Hamiltonian $H_{\rm
int}^{(0)}$ commutes with itself at different times and hence we can
write the two-time influence functional as
\begin{equation}
\rho_M= \langle n_i | e^{ i\int_0^{\tilde T}
dtH_{\rm int}^{(0)}(\tilde r(t))/\hbar} 
e^{-i\int_0^T dtH_{\rm int}^{(0)}(r(t))/\hbar} |n_i \rangle.
\end{equation}
Since the exponents of the two operators in the influence functional 
commute, $\rho_M$ becomes the matrix element of an SU($6$) transformation 
between SU($6$) basis states, in other words it is a representation matrix 
element for this group and can be easily calculated using standard 
techniques.  The two-time influence functional for the $sd$-version of IBM 
was calculated in Ref.~\cite{ibm2} and, for the particular case of the 
SU(3) limit, in Ref.~\cite{ibm1}.

The above method corresponds to a linear coupling approximation.  Due to the 
exponential nature of fusion cross sections, it is not clear at the outset 
whether one can include the effects of the higher order terms in $H_{\rm 
int}$ through a renormalization of the coupling strength.  We next discuss 
an alternative method, whereby one can include the effects of coupling to all 
orders in the interaction Hamiltonian.  In this case the solution of 
Eq.~(3) is no longer an element of the appropriate group.  Nevertheless, we 
can exploit the symmetry properties of the resolvent operator directly 
without utilizing its path integral representation.  Such a Green's 
function approach has also been used to study quantum tunneling in a heat 
bath \cite{yoram}.

To include the effects of couplings to all orders, the interaction
Hamiltonian in Eq.~(1) is written as
\begin{equation}
H_{\rm int}({\bf r}, \xi) = V_{\rm Coul}({\bf r}, \xi) +
V_{\rm  nuc}({\bf r}, \xi).
\end{equation}
Here the Coulomb part is
\begin{eqnarray}
V_{\rm Coul}({\bf r}, \xi)&=&\frac{Z_1Z_2e^2}{r}\left(1 +
\frac{3}{5}\frac{R_1^2}{r^2}\hat O\right) \quad (r>R_1), \nonumber\\
&=&\frac{Z_1Z_2e^2}{r} \left(1 +  \frac{3}{5} \frac{r^2}{R_1^2}
\hat O\right) \quad (r<R_1),
\end{eqnarray}
and the nuclear part is taken to have the Woods-Saxon form,
\begin{equation}
V_{\rm nuc}({\bf r}, \xi) = -V_0\left(1 + \exp{r -R_0 -  R_1
{\hat O}(\hat{\bf r}, \xi)\over a}\right)^{-1}. 
\end{equation}
In Eqs.~(16) and (17), $R_0$ is the sum of the target and projectile
radii and $R_1$ is the mean radius of the deformed target. $\hat O$ is
a general coupling operator between the internal coordinates and the
relative motion which we take to be
\begin{equation}
\hat O(\hat{\bf r}, \xi) = \sum_ {k} v_k T^{(k)}(\xi) \cdot 
Y^{(k)}(\hat{\bf r}) .
\end{equation}
The $v_k$ represent the strengths of the various multipole transitions in 
the target nucleus.  In the standard IBM with $s$ and $d$ bosons, the 
possible transition operators have $k = 2,4$; $k=0$ and odd-$k$ values 
being excluded for reasons stated after Eq.~(9).  The quadrupole and 
hexadecapole operators are given by
\begin{eqnarray}
T^{(2)}&=&[s^{\dagger}{\tilde d}+d^{\dagger}s]^{(2)} + \chi
[d^{\dagger}{\tilde d}]^{(2)},\\ T^{(4)}&=&[d^{\dagger}{\tilde
d}]^{(4)}.
\end{eqnarray}
We adopt the ``consistent-Q'' formalism of Casten and Warner 
\cite{warnercasten}, in which $\chi$ in Eq.~(19) is taken to be the same as 
in $H_{\rm IBM}$ (fitted to reproduce the energy level scheme and the 
electromagnetic transition rates of the target nucleus), and thus is not a 
free parameter.

To calculate the fusion cross section to all orders, we consider the
resolvent operator for the system
\begin{equation}
G^{+}(E) = \frac{1}{E^{+}-H_k-H_{\rm IBM}({\xi})-H_{\rm int}(r, \hat O)}.
\end{equation}
The basic idea is to find the interaction representation which diagonalizes 
$H_{\rm int}$.  Note that, in general, it is not possible to diagonalize 
$H_{\rm IBM}$ and $H_{\rm int}$ simultaneously, so the effect of the 
excitation energies in the target nucleus are ignored in this approach.  To 
this end, we need to identify the unitary transformation which diagonalizes 
the operator $\hat O$
\begin{equation}
\hat O_d = {\cal U} \hat O {\cal U}^{\dagger},
\end{equation}
so that we can calculate its eigenvalues and eigenfunctions
\begin{equation}
\hat O_d | n \rangle = h_n | n \rangle .
\end{equation}
Assuming the completeness of these eigenfunctions
\begin{equation}
\sum_n | n \rangle \langle n| =1,
\end{equation}
one can write the matrix element of the resolvent as
\begin{eqnarray}
\langle \xi_f , r_f \> | \>  G^{+}(E) \> | \> \xi_i , r_i \> \rangle
 &=& \langle \xi_f , r_f | {\cal U}^{\dagger} {\cal U}
\left[  E^{+}-H_k-H_{\rm int}(r, \hat O) \right]^{-1} {\cal
U}^{\dagger} {\cal U} | \xi_i , r_i \rangle, \nonumber\\  
& = & \langle
\xi_f , r_f | {\cal U}^{\dagger} \left[  E^{+}-H_k-H_{\rm int}(r, \hat
O_d) \right]^{-1}  \sum_n |  n \rangle \langle n| {\cal U} | \xi_i ,
r_i \rangle, \nonumber\\ 
& = & \sum_n \langle \xi_f | {\cal
U}^{\dagger} | n \rangle \langle n|  {\cal U} | \xi_i \rangle \langle
r_f | G^{+}_n | r_i \rangle,
\end{eqnarray}
where
\begin{equation}
G^{+}_n(E) = \frac{1}{E^{+}-H_k-H_{\rm int}(r, h_n)}.
\end{equation}
$G^{+}_n(E)$ in Eq.~(26) is the resolvent operator for one-dimensional 
motion in the potential $H_{\rm int}(r, h_n)$, whose fusion cross section 
can be easily calculated within the standard WKB approximation.  The total 
cross section follows from multiplying these eigenchannel cross sections by 
the weight factors indicated in Eq.~(25).  The calculation of the weight 
factors $\langle n| {\cal U} | \xi_i \rangle$ within the IBM is 
straightforward and is given in Refs.~\cite{ibm3,ibm5}, to which the 
reader is referred for further details.  

Comparison of the linear and all orders coupling calculations shows that 
both can describe fusion cross sections equally well.  However there are 
important differences between the two approaches when one considers barrier 
distributions, and only through the latter, can one obtain a good 
representation of the data \cite{ibm3}.  Another advantage of the all 
orders coupling is that the strengths used are very close to the 
deformation parameters and it is easier to establish a contact with the 
geometrical model.  (In contrast, rather large strengths are required in 
the linear coupling approximation.)

\section{A Systematic Study of Subbarrier Fusion in Rare-Earth Nuclei}

\indent 

We carried out a systematic study of subbarrier fusion data,
accumulated over the last few years, for rare-earth nuclei
\cite{ibm4}.  To illustrate the quality of this global fit, in Fig.~1
we compare our theoretical result for the $^{16}$O + $^{148}$Sm,
$^{16}$O + $^{154}$Sm, and $^{16}$O + $^{186}$W systems with the data
from the Australian National University.  There is excellent agreement
between the IBM-based model and the data, as the target nucleus
changes from vibrational ($^{148}$Sm) to deformed with positive
($^{154}$Sm) and negative ($^{186}$W) hexadecapole moments.

Having extracted a consistent set of parameters, we can predict fusion
cross sections and barrier distributions in other rare-earth nuclei.
In this respect, the transitional Os-Pt region is of particular
interest.  One expects that barrier distributions for transitional
nuclei exhibit sharp changes due to shape-phase transitions, contrary
to vibrational and rotational nuclei where the cross sections increase
smoothly with increasing deformation or mass number \cite{ibm4}.  An
accurate measurement of subbarrier fusion cross sections for
transitional nuclei such as Pt and Os provides a sensitive test for
competing models in this region and possibly point to new directions
in research. A subsequent measurement of the fusion cross sections for
$^{40}$C+$+^{194}$Pt and $^{40}$C+$+^{194}$Os systems confirmed 
our predictions \cite{bierman}.

It is also possible to generalize the previous formalism to include
arbitrary kinds of bosons in the target nucleus and investigate
whether $g$ bosons have any discernible effects on subbarrier fusion
reactions.  We found that \cite{ibm5} except for slight differences in
the barrier distributions (which can be made even smaller by fine
tuning the coupling strengths), there are no visible differences
between the {\it sd} and {\it sdg} model predictions. The similarity
of the results implies that subbarrier fusion probes the overall
coupling strength in nuclei, but otherwise is not sensitive to the
details of the nuclear wavefunctions. In this sense, subbarrier fusion
is in the same category as other static quantities (energy levels,
electromagnetic transition rates), and does not seem to constitute a
dynamic probe of nuclei, in contrast to proton scattering \cite{proton}.

Another experimental test of our model would be to study angular
momentum  distributions in subbarrier fusion, which can be determined
reasonably  accurately from the gamma-ray multiplicities data or from
the measurement of relative populations of the ground and isomeric
states in evaporation residues \cite{van}. Vandenbosch and his
collaborators at the University of Washington, Seattle devoted
considerable time to the measurement of angular momentum distributions
\cite{van,sea}. In Fig.~2, we present angular momentum distributions
calculated using our model for a number of systems \cite{ibm6}.  We
should emphasize that these angular momentum distributions are not
direct experimental observables. However there are relations of
rather general nature  relating angular momentum distributions to the
moments of fusion  cross sections \cite{nonibm} which can be used  to
assess the consistency of the direct measurements of the fusion cross
sections with the angular momentum distributions extracted from
gamma-ray multiplicities.

We would like express our gratitude to our collaborators, J.  Bennett,
A.J.   DeWeerd, and N.  Takigawa, without whom the work reported here
would not be  possible.  This research was supported in part by the
U.S.  National  Science Foundation Grants No.  PHY-9605140 and
INT-9315876, in  part by the University of Wisconsin Research
Committee with funds provided  by the Wisconsin Alumni Research
Foundation, and in part by grants from the Australian Research Council 
and the Department of Industry, Science and Technology.

{}

\centerline{\bf Figure Captions}

Fig.~1.  Comparison of fusion cross sections and barrier distributions 
predicted using the Interacting Boson Model with the data as described in 
the text.  The parameters are given in Ref.~\cite{ibm4}.

Fig.~2.  Comparison of angular momentum  distributions  predicted
using the Interacting Boson Model with the data.   The  parameters are
given in Ref.~\cite{ibm6}.

\end{document}